\documentclass[12pt]{article}

\usepackage{amsmath}
\usepackage{amsfonts}
\newcommand{\lag}{{\mathcal L}}

\title{Decaying Cosmological Constant of the Inflating Branes in the Randall-Sundrum -Oda Model}
\author{Laura Mersini\\Department of Physics\\University of Wisconsin-Milwaukee\\Milwaukee, WI 53201\\lmersini@uwm.edu\\Wisc-Milw-99Th-14}
\date{September 20, 1999}
\begin{document}
\maketitle
\begin{abstract}
We examine the issue of the cosmological constant in the $many$ $inflating$ branes scenario, extending on two recent models  by I.Oda and Randall-Sundrum.The exact solution in a closed form is found in the slow roll approximation of the radion. Defining an effective expansion rate, which depends on the location of each brane in the fifth dimension and demanding stability for this case we show that each positive tension brane has a localized, decaying cosmological constant (the opposite process applies to the negative energy branes [4]) .\\
The reason is that the square of the effective expansion rate enters as a source term in the Einstein equations for the branes.Thus the brane has  two scale factors depending on time and the fifth dimnesion respectively .The brane will roll along the fifth dimension in order to readjust its effective expansion rate in such a way that it compensates for its internal energy changes due to inflation and possible phase transitions. 
\end{abstract}
\pagebreak

\section{Setup}

In this paper we consider and find solutions to a five dimensional model with many inflating branes along the fifth compact direction, thus extending on the recent models proposed by Randall-Sundrum and Oda [1,2,3]. 
Each brane is taken as a point in the fifth dimension (negligible thickness) $S^1$. They act as localized cosmological constants or alternatively as gravitational point sources.The whole 5D setup should be static in order to preserve Poincare invariance. That means that the 5D cosmological constant $\Lambda_5$ should cancel out the collected contribution of the 4D cosmological constants $\Lambda_{4i}$ of each brane (weighted by the appropriate conformal factor as shown below in Sect.2). Note that each $\Lambda_{4i}$ depends on the position of the i-th brane in $S^1$.

\noindent The existence of solution to the Einstein equation in the many branes
scenario does not allow for a singular orbifold geometry $S^1/Z_2$ but
instead, it requires a smooth manifold $S^1$ [1].

\noindent The 5D metric ansatz considered here is:
\begin{equation}
ds^2 = g_{MN} dx^M dx^N = u(z)^2 dt^2 - a(z,t)^2 dx^2 - b(z)^2 dz^2
\end{equation}
with $z$ ranging between $0$ and $2 L$.
 
\noindent The 5D Einstein-Hilbert action is:
\begin{equation}
S = \frac{1}{2\kappa^2_{5}} \int d^4 x \int^{2L}_0 dz \sqrt{-g} (R+2\Lambda_5)
+ \sum^n_{i=1} \int_{z=L_i} d^4 x \sqrt{-g_i}({\cal L}_i + V_i)
\end{equation}
where the 4D metric $g_{i\mu\nu}$ is obtained from the 5D one as follows:
$g_{i\mu\nu}({\bf x},t) = g_{MN}({\bf x},t,z = L_i)$ and ${\cal L}_i, V_i$
are the lagrangian of the matter fields and the vacuum energy respectively,
in the i-th brane.The 5D gravitational coupling constant $\kappa_{5}$ is related to the 4D Newton constant $G_N$ through $\kappa_5 =16 G_{N}\pi \int dz\sqrt{-g_{55}} =16 G_{N}\pi L_{phys} $.
\vspace*{0.5cm}\\
\section{Many Inflating Branes Scenario. Solution}

We are interested in constructing a solution for the inflating branes. Thus
in Eqn. (1) we take:
\begin{equation}
\begin{array}{ccl}
a(z,t) & = & f(z)  v(t)\\
b(z) & = & f(z)\\ 
u(z) & = & f(z) 
\end{array}
\end{equation}
The dynamics of the boundaries is based on the slow roll assumption.The energy density of each brane is denoted by ${\cal L}_i$ where we have separated a constant vacuum energy contribution, $V_{i}$. The following equations result from varying the action in Eqn.(2) with
respect to the metric, [1,2,3]:
\begin{eqnarray}
\frac{1}{f^2}\left[ (\frac{\dot{v}}{v})^2 \right] - \frac{1}{f^2}
\left[\frac{f''}{f} + (\frac{f'}{f})^2 \right] & = &
- \frac{\kappa^2}{3 f}[\sum_i({\cal L}_i + V_i)\delta(z - L_i)] + 
\frac{\Lambda_5}{3}\nonumber \\
& & \nonumber\\
\frac{1}{f^2}\left[ (2\frac{\ddot{v}}{v})+ (\frac{\dot{v}}{v})^2\right] 
- \frac{3}{f^2}\left[ \frac{f''}{f} + (\frac{f'}{f})^2\right] & = &
- \frac{\kappa^2}{f}\left[ \sum_i({\cal L}_i + V_i)\delta(z - L_i)\right]
+{\Lambda_{5}}\nonumber\\
& & \nonumber\\
\frac{1}{f^2}\left[ (\frac{2\ddot{v}}{v})+ (\frac{\dot{v}}{v})^2\right]
 - \frac{2}{f^2}(\frac{f'}{f})^2 & = & \frac{\Lambda_5}{3} 
\end{eqnarray}
where prime and dot denote differentiation with respect to z and t
respectively.  It is straightforward from the Eqns. (4) above (as well as  the component $G_{55}=0$), that $v(t)$
satisfies:
\[ \frac{\ddot{v}}{v} = \frac{\dot{v}^2}{v} \] 
which gives for $v$:
\begin{equation}
v(t) = v(0)exp[Ht]
\end{equation} 
Thus Eqns. (4) become:
\begin{equation}
\begin{array}{lcc}
\frac{H^2}{f^2} - \frac{1}{f^2} \left[ \frac{f''}{f} + 
                  (\frac{f'}{f})^2 \right] & = & 
- \frac{\kappa^2}{3f} [\sum_i({\cal L}_i + V_i)]\delta(z - L_i) + 
\frac{\Lambda_5}{3} \\
& & \\
\frac{H^2}{f^2} - \frac{2}{f^2}(\frac{f'}{f})^2 
 =  \frac{\Lambda_5}{6}
\end{array}
\end{equation}
The effective expansion rate is defined as $H_{eff} = H / f(z) $ .  
The effective expansion rate of each brane, located at $z = L_i$ in the
extra dimension, is then given by $H_{ieff} = H / f(L_i)$.  Thus each brane,
 depending on its position in the fifth dimension, sees a different 
expansion rate resulting from the fact that  the canonical proper time has the following scaling on the position of the brane, $d \tau_{i} = f( L_{i}) dt$. Denote the absolute value of $\Lambda_5$ by $\Lambda$. Since we consider an  $ADS_5$ 
geometry, $\Lambda$ is positive and satisfies the following relation:
\[ \Lambda = - \Lambda_5 .\]
We will find the solution by construction. 
Assume the ansatz in Eqn. (7) for the warp factor $f(z)$. Then its derivatives are given by Eqns. (8,9).
\begin{eqnarray}
f(z)   =  {\alpha}
{\displaystyle Sinh[g(z)]^{-1} }
& & \\           
f'( z) =  -{\alpha} \displaystyle \frac {Cosh [g(z)]}{ Sinh[g(z)]^{2} } g'(z)
& & \\
f''(z) = -{\alpha }       (\displaystyle \frac{ g'(z)^2 }{ Sinh[g(z)] } -           - \frac { 2 Cosh[g(z)]^{2} }{ Sinh[g(z)]^{3} }  {g'}^{2} +
\frac { Cosh[g] }{ Sinh[g]^2 } g'')
\end{eqnarray}
The function $g(z)$ [1,3], describing even branes (domain walls) of positive and negative energy positioned  along $S^1$ is of the following form
\begin{eqnarray}
f(z) &  =\alpha \displaystyle{Sinh[g(z]^{-1}}             \\
g(z)  &=\left(\displaystyle{\sum^{n-1}_{i=1}}(-1)^{i+1}|z-L_i|+L\right)(-\beta)\\
g''(z)&=2\left(\displaystyle{\sum^{n-1}_{i=1}}(-1)^{i+1} 
            \delta |z - L_i| \right) (-\beta) 
\end{eqnarray}
The even $(n-1)$ branes are located at
$L_i (i = 2, \dots n-1)$.
One always ends up with even branes as it is clear from topological 
consideration.  Each brane is a pointlike gravitational source in the
fifth dimension,  i.e. there are flux lines extending from one brane to
the others (mutual interaction).  Those flux lines should close in order
to preserve the stability of the 5th dimension (see [4]).  Thus each positive energy brane in
$S^1$ is alternating with its counterpart (negative tension brane) so that the number of positive
energy branes equals the number of negative energy branes.\\
Replacing the above Eqns. (7-12) for $f(z)$ in the Einstein Eqns.(6) we get the following relation
\begin{eqnarray}
\nonumber (H^2 - {g'}^2)+ \frac {\Lambda /3  - 2{g'}^2}{ \displaystyle Sinh[g]^2  }
&= & \frac{( \sum_i - \kappa_{5} ({\cal L}_i + V_i) - g''
 {\displaystyle Cosh[g(L_i)]}) \delta(z - L_i)}{Sinh[g]} \\
\nonumber (H^2 - {g'}^2)  & =&  - \displaystyle \frac{\Lambda \alpha^2}{6 Sinh[g]^2}  + 
\frac { {g'}^2 }{ Sinh[g]^2 } 
\end{eqnarray}
There is a  solution to Eqns. (9) only if $\beta = H$ and 
$\alpha^2 = \frac {6 H^2}{\Lambda}$ in Eqns. (10-12).The value of $ H $ is decided by $\Lambda$ and $ L_{phys}$.
The conditions on the energy densities of the branes and their respective effective expansion rates $H_{ieff}$ become:
\begin{eqnarray}
{\cal L}_i +V_i   &=& \frac {2}{\kappa_5} (-1)^{(i+1)} \sqrt{ \frac {\Lambda}{6} }(\displaystyle Cosh[g(L_i)]) \\
H_{ieff} &=& \sqrt{ \frac {\Lambda}{6} } (\displaystyle  Sinh[g(L_i)])
\end{eqnarray}   
The warp factor $f(z)$ found in Eqns. (10-12) satisfies Einstein equations and has a minimum at $z=0, 2L$ and a maximum at $z= L$. (For subtleties related to ensuring the periodicity of $f(z)$ and the overlap of branes located at $z=0, 2L$ see [1]. The physical length, $L_{phys} = 2 L$ calculated by the formula in Sect.1, depends on the energies of the branes as they are the gravitational sources causing the curvature of the fifth dimension. It is straightforward to  perform the calculation explicitly ,see [3]). 
It is clear from Eqns. (14, 15) that branes positioned near the minimum of the warp factor, $z=0$ (the 'TeV branes') in the fifth dimension, have the maximum energy and effective expansion rate allocated to them while the ones near the position of the  maximum warp factor $z=L_{phys}$ ('Planck branes'), are almost empty of energy and have the lowest effective expansion rate ($H_{ieff}$  at $z=L_{phys}$ is almost but not identically zero, a result also found by [4,10]).
Thus we end up with the following relation between the expansion rate and the energy density for the branes
\begin{equation}
|\frac {\kappa_5 ( {\cal L}_i + V_i)}{2}|^2 - H_{ieff}^2 = \frac {\Lambda}{6}
\end{equation}
Notice the unusual 'Friedman relation' between the energy density and the expansion rate for each individual  brane [4,5,10]. An important result of the Eqn. (16) is that each brane has a different expansion rate depending on its position in $S^1$. Clearly from (16), branes located around the position of  minimum warp factor have maximum expansion rate, while the 'Planck branes' have nearly zero expansion rate and energy. There is a deep physical reason behind the relation of Eqn. (16). It is related to the induced radion potential and the low suppression of its modes to the TeV branes and high supression to th e Planck branes. We outline this reasoning below, in Sect.3, (however a more complete and detailed treatment ofthe relation of  radion dynamics and inflating branes will be reported soon in [12]). 

\section{ Discussion}

In order to gain insight in the relation of Eqn. (16) above, we would like to take the point of view of Dvali et al.[11] that although the inflaton is a brane mode in the ground state, it behaves as an 'inter-brane' mode that describes a relative separation of the branes in the extra dimension. This approach clarifies the physics of brane dynamics during inflation. It is the radion (inter-brane separation) which in the effective 4D description becomes the inflaton in the branes. The 'TeV branes'' have minimum suppression of the radion modes as compared to the 'Planck branes', thus they will have maximum inflation  which is confirmed by Eqn. (16). The coupling of 'Planck branes' have maximum suppression to the radion, hence nearly zero inflation rate since the the energy dumped into them by the radion is nearly zero. Branes inflate with different rates for the above reason. Their coupling to the radion depends on their position/angle in the $S^1$. 
We have avoided introducing a radion potential and assumed a slow-roll behaviour for the branes. Hence this potential is implicitly there as it is induced by the branes. The slow-roll holds for the most part except around $z=L$ where our analysis would break down. For a complete treatment we need to consider the full dynamics of the branes, radion stability and bulk fields. These will provide a  mechanism whereby branes roll towards decreasing values of $H_{eff}$ (also decreasing the effective 4D cosmological constant) thus exiting inflation, reheating the universe and recovering the usual Friedman relation in late cosmology (work in progress [12]).   
It is always useful to use analogies of the system under consideration with the condensed matter systems. The above scenario is very similar to a string of molecules confined to reside in a circle. Each brane would represent a molecule with four internal degrees of freedom (branes are 4D worlds), chemical potential (gibbs free energy per molecule) $ H_{ieff}^2$, energy $\lag_i + V_i$, arranged in such a configuration that the total system is in equlibrium.For the moment let's concentrate on the positive tension branes (the opposite will go for the negative energy ones). Since each brane inflates, the energy $\lag_i$ of the matter fields and the temperature tend to change towards decreasing values. The vacuum energy $V_i$ of each brane can also change towards decreasing values due to internal phase transitions driving the vacuum to a lower minimum. In our analogy, $V_i$ would be the analog of the latent heat of each molecule. But each position in the circle $S^1$ belongs to a certain $H_{ieff}$  energy level ('molecules' with different 'chemical potential'). Thus, when the total internal energy of the brane decreases (for the reasons mentioned above) the brane moves to the next position in the fifth dimension thus changing its $H_{ieff}$ in order to accommodate (compensate for) the internal energy changes and to preserve the equlibrium configuration of the system as a whole. Each brane has the freedom to do the readjustment through rolling due to the fact that the  effective expansion rate enters as a constant source term in the 4D brane equations but it is a function of the fifth dimension, $H_{eff} = Hf(z)^{-1}$.\\
Therefore, each positive (negative) energy brane will move towards (away from) the lowest $H_{eff}$  (i.e. maximum (minimum) $f(z)$) to compensate for internal energy changes. In short  $H_{eff}$ will $decay$ nearly $exponentially$ , (see Eqn.(15) for the exponential dependence of $f(z)$), by rolling down $f(z)^{-1}$. Ultimately all positive energy branes will locate at the $f(z)$ maximum point  and the negative energy ones at the $f(z)$ minimum point.\\
As the internal 4D world of each brane is not aware of the fifth dimension z , $H_{eff}^2$  plays the role of a decaying 4D cosmological constant to each brane. The reason, as mentioned, is that when $\lag_i$ (which is a dynamic quantity) changes internally for each brane, then $H_{ieff}(z)^2$ ( which reflects the coupling of each 4D brane to the 5D gravity) is forced to change by displacing to a new position such that $(\lag_i + V_i) -\alpha H_{ieff}^2$ cancels out the flux lines/ energy allocated by $\Lambda_5$ to that position in $S^1$. (The constant $\alpha=\frac{3}{\kappa^2 }$). Obviously the mass scales and proper time $\tau_i$ of each brane also change while the brane changes position in $S^1$, thus a set of hierarchy scales. \\   
To put it differently, there are two competing scale factors or redshifts affecting each brane, the $z$ and  $t$ factors. A redshift in $t$ forces a 'blueshift' in $z$ because of the requirement of equlibrium configuration of the system as a whole (pictorially ,the flux line should be closed in this static 5D model. Since branes interact gravitationally with each other along the $S^1$, the flux lines originating from positive tension branes should end up in the negative tension branes. This is also consistent with the solution in Sect.3 that, in each pair of branes, while the positive energy branes shifts towards maximum $f(z)$ its counterpart (negative energy) shifts towards minimum $f(z)$ such that the flux lines remain close).\\
There remains an open issue,namely: What happens to the branes as,ultimately,they all pile up to only two point of $S^1$, (stability issues are discussed in detail in [4]) , the positive tension ones at maximum $f(z)$ and negative tension branes at minimum $f(z)$. Naive energy arguments would suggest they recombine into a single 4D brane in either point (a chemical equlibrium situation), thus the problem would reduce to two single branes as in [2,3]. However this isssue remains to be investigated more closely [12].\\
In conclusion, in this many inflating branes scenario, the equlibrium condition of the 5D model, together with inflation and phase transitions taking place internally in each 4D brane, drives a decaying cosmological constant,which while it may not depend on the internal 4 dimensions, does depend on the fifth one and thus has the freedom to rearrange thereby preserving the equlibrium configuration.
\noindent\\
Acknowledgment:
I would like to thank Prof. L. Parker for our regular discussions.This work was supported in part by NSF Grant No. Phy-9507740.
\pagebreak

{\large{\bf References}}
\begin{itemize}
\item[{1}]  Ichiro Oda,$hep-th/9908076$; Ichiro Oda, $hep-th/9909048$
\item[{2}]  L. Randall and R.Sundrum, 'A large mass hierarchy from a small extra dimension', $hep-ph/9905221$
\item[{3}]  T.Nihei, $hep-ph/9905487$
\item[{4}]  P.Steinhardt, $hep-ph/9907080$; L.Mersini, $gr-qc/9906106$; C.Csaki et al, $hep-ph/990651$; W.D.Goldberger and M.B.Wise, $hep-ph/9907447, hep-ph/9907218$   
\item[{5}]   A.Lukas, B.A.Ovrut and D.Waldram, $hep-th/9902071$
\item[{6}]   N.Kaloper and A.Linde, $hep-th/9811141$
\item[{7}]   J. Lykken and L.Randall, $hep-th/9908076$
\item[{8}]   N. Arkani-Hamed, S, Dimopoulos, G.Dvali and N.Kaloper, $hep-th/9907209$
\item[{9}]   N. Verlinde, $hep-th/9906182$
\item[{10}]  C. Csaki et al $hep-ph/9906513, hep-th/9908186$, J.Cline et al $hep-ph/9906523$, J. Cline $hep-ph/9904495$ 
\item[{11}]  G. Dvali and Henry Tye, $hep-ph/9812483$, Daniel J.H. Chung, Edward W. Kolb and Antonio Riotto, $hep-ph/9802238$
\item[{12}]  P. Binetruy et al $hep-th/9905012$, L. Mersini, 'Radion potential and Brane Dynamics',  $in$ $ preparation$ 
\end{itemize}]
\end{document}